\documentclass[11pt]{article}%
\usepackage{moriond,epsfig}
\usepackage{amsmath}
\usepackage{amsfonts}
\usepackage{amssymb}
\usepackage{graphicx}%
\def\be{\begin{equation}}
\def\ee{\end{equation}}
\def\bea{\begin{eqnarray}}
\def\eea{\end{eqnarray}}

\begin{document}

\title{ ELECTROWEAK PHYSICS\ FROM\ NUTEV}
\author{T. Bolton \ \textit{(for the NuTeV Collaboration)}\\Kansas State University, Manhattan, KS 66502 USA}
\maketitle

\abstracts{
NuTeV has performed precise measurements of neutral-current to
charged-current cross section ratios using intense high energy neutrino and anti-neutrino
beams on a primarily steel target at the Fermilab Tevatron.  A null hypothesis test of the
standard model allows the extraction
$\sin^2\theta_W^{\nu N}(\equiv 1-M_W^2/M_Z^2)=0.2277\pm0.0013(stat)\pm0.0009(syst)$,
a value that differs from predictions of global electroweak fits by $+3.0\sigma$.
}

\section{Background}

High energy neutrino and antineutrino beams scattered from an isoscalar target
$N$ allow measurement of two cross section ratios that can be compared to
robust electroweak predictions\cite{LS} at moderate space-like momentum
transfer:%
\begin{equation}
R^{\nu}\left(  R^{\bar{\nu}}\right)  =\frac{\sigma\left(  \nu_{\mu}\left(
\bar{\nu}_{\mu}\right)  N\rightarrow\nu_{\mu}\left(  \bar{\nu}_{\mu}\right)
X\right)  }{\sigma\left(  \nu_{\mu}\left(  \bar{\nu}_{\mu}\right)
N\rightarrow\mu^{-}\left(  \mu^{+}\right)  X\right)  }=g_{L}^{2}+r\left(
\frac{1}{r}\right)  g_{R}^{2}, \label{rnudef}%
\end{equation}
with $g_{L,R}^{2}=g_{L,R}^{2}\left(  u\right)  +g_{L,R}^{2}\left(  d\right)
$, $r=\frac{\sigma\left(  \bar{\nu}_{\mu}N\rightarrow\mu^{+}X\right)  }%
{\sigma\left(  \nu_{\mu}N\rightarrow\mu^{-}X\right)  }\simeq0.5$, \ and, at
tree level in the standard model, $g_{L}^{2}=\frac{1}{2}-\sin^{2}\theta
_{W}^{\nu N}+\frac{5}{9}\sin^{4}\theta_{W}^{\nu N}$, $g_{R}^{2}=\frac{5}%
{9}\sin^{4}\theta_{W}^{\nu N}.$ The combination\cite{PW}%
\begin{equation}
R^{-}=\frac{R^{\nu}-rR^{\bar{\nu}}}{1-r}=\frac{\sigma\left(  \nu_{\mu
}N\rightarrow\nu_{\mu}X\right)  -\sigma\left(  \bar{\nu}_{\mu}N\rightarrow
\bar{\nu}_{\mu}X\right)  }{\sigma\left(  \nu_{\mu}N\rightarrow\mu^{-}X\right)
-\sigma\left(  \bar{\nu}_{\mu}N\rightarrow\mu^{+}X\right)  },
\label{rminusdef}%
\end{equation}
is independent of strong interaction contributions and equal, in leading
order, to $R^{-}=\frac{1}{2}-\sin^{2}\theta_{W}^{\nu N}.$

Many experimental corrections\cite{nutev prl} must be applied to produce the
ratios in Eq. \ref{rnudef}, and significant QCD corrections are needed to test
the coupling predictions. Most notable of the latter category is the
correction for charm production necessitated by the kinematic suppression
associated with the charm mass. Uncertainties in the implementation of this
correction limited the best experiment previous to NuTeV\cite{ccfr} to a
precision $\Delta\sin^{2}\theta_{W}=0.0041$, corresponding to an equivalent
$W$ mass uncertainty of $210$ MeV$/c^{2}.$ NuTeV constructed sign-selected
neutrino beams with sufficient intensity and purity to effectively extract
$R^{-}$. \ Corrections for charm production needed in passing from Eq.
\ref{rminusdef} to $\sin^{2}\theta_{W}^{\nu N}$ still exist, but at a
considerably reduced level because they are CKM-suppressed and dependent upon
only high $x$ ($\approx$ high $\nu-$quark mass) valence quark distributions.

\section{NuTeV Results and Implications}

\bigskip Assuming the standard model, which allows for a calculation of
$R^{-}$ in terms of $\alpha_{EM}$, $G_{F}$, $M_{Z}$, $M_{\text{top}}$, and
$M_{\text{Higgs}}$, NuTeV finds%
\begin{align}
\sin^{2}\theta_{W}^{\nu N}\left(  \equiv1-M_{W}^{2}/M_{Z}^{2}\right)   &
=0.22773\pm0.00135\text{(stat)}\pm0.00093\text{(syst)}\label{nutev result}\\
&  -0.00022\times\left(  \frac{M_{\text{top}}^{2}-\left(  175\text{ GeV}%
/c^{2}\right)  ^{2}}{\left(  50\text{ GeV}/c^{2}\right)  ^{2}}\right)
+0.00032\times\ln\left(  \frac{M_{\text{Higgs}}}{150\text{ GeV}}\right)
,\nonumber
\end{align}
where \textquotedblleft$\equiv1-M_{W}^{2}/M_{Z}^{2}$\textquotedblright%
\ denotes a choice of the on-shell scheme for radiative
corrections\cite{bardin-1,bardin-2} that relates $\sin^{2}\theta_{W}^{\nu N}$
directly to the physical gauge boson masses. Taking $M_{\text{top}}=175$
GeV$/c^{2}$ and $M_{\text{Higgs}}=150$ GeV and using the precisely measured
$Z^{0}$ mass, the NuTeV measurement implies $M_{W}=80.14\pm0.08$ GeV$/c^{2}$.
The overall $\Delta\sin^{2}\theta_{W}^{\nu N}$ betters the previous neutrino
world average by a factor of two and is statistics-dominated. \ The
uncertainty in $M_{W}$ compares favorably to that obtained from direct
extractions and other precision electroweak measurements.

NuTeV also relaxes standard model assumptions and obtains the
couplings\footnote{These numbers have been updated to correct a small
numerical error in the NuTeV\ publication.} $g_{L}^{2}=0.30005\pm0.00137$,
$g_{R}^{2}=0.03076\pm0.00110$ by omitting electroweak corrections save for the
large and experiment-dependent QED parts that approximately factor. Results
for $g_{L,R}^{2}$ have stronger dependences on the neutrino charm production
model and are more likely to be affected by higher order QCD corrections than
that for $\sin^{2}\theta_{W}$\cite{davidson}.

\subsection{Experimental Details}

A description of NuTeV analysis details is available elsewhere\cite{nutev
prl}. \ 

The largest experimental uncertainty, besides statistics, is associated with
imperfect knowledge of the $\sim1.7\%$ level $\nu_{e}/\bar{\nu}_{e}$
background flux $\left(  \Delta\sin^{2}\theta_{W}^{\nu N}=0.00039\right)  $.
\ The NuTeV beamline suppresses relatively poorly constrained neutral hadron
sources of this flux, leaving charged kaon decays as the dominant source. This
contribution is in turn tightly constrained by the observed $\nu_{\mu}%
/\bar{\nu}_{\mu}$ flux produced by $K^{\pm}$ in charged current event samples.
Charm particle decays in the neutrino production target produce the next
largest $\nu_{e}/\bar{\nu}_{e}$ flux contribution; this source is constrained
by measurement of \textquotedblleft wrong sign\textquotedblright\ charged
current event rates in the experiment\cite{drew prd}.

The largest model uncertainty in the $\sin^{2}\theta_{W}^{\nu N}$ extraction
arises from residual charged current charm production, $\left(  \Delta\sin
^{2}\theta_{W}^{\nu N}=0.00047\right)  .$ The magnitude of this term has been
verified by others\cite{davidson}. Its computed size is independent of the
details of the charm production model for the $\sin^{2}\theta_{W}^{\nu N}$
extraction from $R^{-}.$

\subsection{Comparison to Other Electroweak Measurements}

A global fit to all electroweak data except neutrino
measurements\cite{lepewwg} implies $\sin^{2}\theta_{W}=0.2227$, $g_{L}%
^{2}=0.3042$, and $g_{R}^{2}=0.0301$, with negligible errors compared to the
NuTeV measurements. The average of direct $W-$mass measurements is
$M_{W}=80.45\pm0.04$ GeV$/c^{2}.$ The NuTeV result is three standard
deviations higher(lower) than predictions for $\sin^{2}\theta_{W}$($M_{W}$),
while $g_{L}\left(  g_{R}\right)  $ are shifted down(up) compared to
predictions. As a consequence, \ $R^{\nu}$, $R^{\bar{\nu}}$, and $R^{-}$ are
all lower than predicted. \ The global electroweak fit without the NuTeV
measurement has $\chi^{2}/N=19.6/14$ ($14\%$ probability); with the NuTeV
result this becomes $\chi^{2}/N=28.8/15$ ($1.7\%$ probability). $\ $%
Essentially all of the $\chi^{2}$ contribution that is greater than $N$ comes
from NuTeV and $A_{fb}^{0,b}$, the forward-backward asymmetry for $b-$quarks
measured at the $Z^{0}$ pole; these are the only two measurements that prefer
a large Higgs boson mass in the global fit. Without $A_{fb}^{0,b}$ and
$\sin^{2}\theta_{W}^{\nu N}$, the global fit prediction for $M_{\text{Higgs}}$
would sink to $\symbol{126}55$ GeV$/c^{2}$, uncomfortably below the direct
search exclusion limit-- though NuTeV's sensitivity for $M_{\text{Higgs}}$ is
minimal(Eq. \ref{nutev result}) and $A_{fb}^{0,b}$ drives the fit.

The statistical situation is, in short, intriguing, but inconclusive. \ It
lies within the bounds of reason to regard the $\sin^{2}\theta_{W}^{\nu N}$
and $A_{fb}^{0,b}$ measurements as simple fluctuations, and to see the overall
global electroweak fit result as yet another ringing endorsement of the
standard model.

\subsection{Possible Standard Model Explanations}

Assuming the NuTeV\ measurement is not a fluctuation, one can consider pursue
\textquotedblleft explanations\textquotedblright\ for the \textquotedblleft
discrepancy\textquotedblright. \ \ Plausible standard model effects that NuTeV
did not explicitely account for in its analysis include nuclear shadowing,
asymmetries in the nucleon strange sea, and nucleon-level isospin violation.

Shadowing can be understood as a very low $Q^{2}$ phenomenon wherein the
exchanged $W^{\pm}$ and $Z^{0}$ bosons fluctuate into vector or axial vector
mesons. Miller and Thomas\cite{miller} argue that shadowing is weaker for
$Z^{0}$ exchange than for $W$ exchange, and that $R^{\nu/\bar{\nu}}$ should
therefor be increased in an iron target compared to simple partonic
expectations, at least for the part of the NuTeV\ data sample with low $Q^{2}%
$. The major problem with their observation is that it has the wrong sign:
NuTeV data show \emph{smaller than expected} $R^{\nu/\bar{\nu}}$. \ One would
also expect minimal shadowing effects in $\sin^{2}\theta_{W}^{\nu N}$
extracted from $R^{-}$ because the vector-meson cross sections are charge
symmetric and cancellations will thus occur in the numerator and denominator
of \ Eq. \ref{rminusdef}.

An asymmetric strange sea $\left(  \bar{s}\neq s\right)  $ can affect
predictions for $R^{-}$ since terms proportional to $s-\bar{s}$ appear in the
numerator and denominator of \ Eq. \ref{rminusdef}. \ The best handle on this
physics comes from a NuTeV analysis\cite{max prd} of the dimuon processes
$\nu_{\mu}/\bar{\nu}_{\mu}N\rightarrow\mu^{\pm}\mu^{\mp}X.$ Dimuon final
states are dominated by charm production, important contributions to which
occur through the charged current sub-processes $\nu_{\mu}s\rightarrow\mu
^{-}c$ and $\bar{\nu}_{\mu}\bar{s}\rightarrow\mu^{+}\bar{c}$. \ NuTeV's
separated beams permit reliable independent extractions of $s$ and $\bar{s}.$
The two distributions are found to be consistent with being equal to one
another, and thus no asymmetry is observed. \ Taking the data at face value
and analyzing it using the same cross section model used to extract $\sin
^{2}\theta_{W}^{\nu N}$ \cite{nutev rapcom}, it is again found that the sign
of the (statistically weak) effect observed using the dimuon samples is
\emph{opposite} that needed to account for the weak mixing angle discrepancy.
NuTeV\ has published its data in a nearly model-independent form that should
allow more detailed examination of these ideas.

Finally, failure to take into account isospin violation can upset the mixture
of $u$ and $d$ quark couplings used in determining $\sin^{2}\theta_{W}^{\nu
N}$. NuTeV's iron target has a $\symbol{126}5.7\%$ excess of neutrons over
protons. This gross effect is accounted for; in fact, computing the
corresponding correction requires NuTeV to make its only significant use of
parton distribution functions not extracted self-consistently from the
experiment itself. \ The more subtle effect not explicitly corrected for
occurs at the nucleon level in the possible breaking of the generally assumed
identities $u_{p}=d_{n}$ and $d_{p}=u_{n}$. \ An early bag model calculation
estimated effects on $\sin^{2}\theta_{W}^{\nu N}$ as large as $0.002$%
\cite{sather}; however more recent calculations\cite{rodionov,cao} yield
estimates of shifts at the $10^{-4}$ level. NuTeV has no ability to probe
nucleon isospin violation directly. \ These effects would have to be inferred
from a global analysis (CTEQ, MRST, GRV...) of deep inelastic scattering and
other experiments that employ proton and deuterium targets. \ 

\subsection{Non-Standard Model Explanations}

The new physics potential of precision neutrino scattering measurements has
long been recognized\cite{langacker}; however, it is challenging to find an
effect that explains the NuTeV\ deviation without contradicting other
precision measurements. \ Davidson \textit{et al.} show that the following
models \emph{do not work}\cite{davidson}: anything generating oblique type
electroweak radiative corrections, models of anomalous neutrino couplings,
extra $Z^{\prime}$ with generation-independent $SU\left(  2\right)  _{L}$
couplings, low energy minimal supersymmetry, and $SU\left(  2\right)  $
singlet or doublet leptoquarks. New physics models they identify that can
explain a significant fraction of the NuTeV\ effect include contact
interactions, possibly mediated by vector leptoquarks at a scale of
$\symbol{126}1.4$ TeV, and a new $U(1)$ $B-3L_{\mu}$ gauge symmetry containing
a $Z^{\prime}$ that de-couples from first generation leptons and mixes weakly
with the standard model $Z$. The new $Z^{\prime}$ is compatible with existing
data if it is either very heavy $\left(  M_{Z^{\prime}}\gtrsim600\text{
GeV}/c^{2}\right)  $ or very light $\left(  M_{Z^{\prime}}\lesssim10\text{
GeV}/c^{2}\right)  $.

Babu and Pati\cite{babu} claim that the NuTeV\ result is predicted by an
extended supersymmetry model with an $SO\left(  10\right)  $ gauge symmetry.
Their model predicts the value of $\left\vert V_{cb}\right\vert $ and the
observed \textquotedblleft neutrino counting deficit\textquotedblright\ at LEP
\ Barshay and Kreyerhoff\cite{barshay} invoke a new parity-conserving neutrino
interaction containing a very heavy new neutral lepton. In this model, the
$\nu_{\mu}$ effectively acquires an internal structure at distances
$\lesssim10^{-18}$ cm. Implications include a $\sin^{2}\theta_{W}^{\nu N}$ in
accord with NuTeV, an accounting for the LEP neutrino deficit, and neutrinos
that acquire strong interaction type cross sections for $E\gtrsim10^{21}$ that
could explain the presence of anomalous ultra-high-energy cosmic ray
interactions \ Giunti and M. Laveder\cite{giunti} attribute the NuTeV effect
to the disappearance of electron-type neutrinos into sterile neutrinos with
oscillation parameters $P\left(  \nu_{e}\rightarrow\nu_{s}\right)
=0.21\pm0.07$ with $\Delta m^{2}=10-100$ eV$^{2}$. However, as noted by
Davidson \textit{et al.}, NuTeV's finding that direct measurement of the
electron neutrino flux agrees with expectations likely already rules this
scenario out. \ NuTeV has recently extended this work into an exclusion region
for $\nu_{\mu}/\bar{\nu}_{\mu}\rightarrow\nu_{e}/\bar{\nu}_{e}$
oscillations\cite{avva}. \ Ma and Roy\cite{ma-1,ma-2} considers two examples
of new gauged $U(1)$ symmetries. The first adds a heavy triplet of new
fermions to each family to provide an alternative see-saw mechanism for
neutrino mass. The second gauges the symmetry $L_{\mu}-L_{\tau}$. Both models
predict TeV scale $Z^{\prime}$ bosons that could explain the NuTeV anomaly if
$ZZ^{\prime}$ mixing is kept small.

In summary, new physics models exist that can explain the NuTeV weak mixing
angle result, but they are not simple extensions of the standard model.

\section*{Acknowledgements}

We thank the staff of the Fermilab Beams, Computing, and Particle Physics
divisions for design, construction, and operational assistance during the
NuTeV experiment. \ This work was supported by the U.S. Department of Energy,
the National Science Foundation, and the Alfred P. Sloan foundation.


\section*{References}

\begin{thebibliography}{99}                                                                                               %


\bibitem {LS}C.H. Llewellyn Smith, Nucl. Phys. \textbf{B}228, 205 (1983).

\bibitem {PW}E.A. Paschos and L. Wolfenstein, Phys. Rev. \textbf{D7}, 91 (1973).

\bibitem {nutev prl}G.P. Zeller, \textit{et al. }(NuTeV Collaboration), Phys.
Rev. Lett. \textbf{88}, 091802 (2002).

\bibitem {ccfr}K.S. McFarland, \textit{et al.} (CCFR Collaboration), Eur.
Phys. J. \textbf{C1}, 509 (1998).

\bibitem {bardin-1}D. Bardin and V.A. Dokuchaeva, JINR-E2-86-260 (1986).

\bibitem {bardin-2}D. Bardin, \textit{et al.}, Comp. Phys. Commun.
\textbf{133}, 229 (2001).

\bibitem {davidson}S. Davidson, \textit{et al.}, hep-ph/0112302, March 2002.

\bibitem {drew prd}A. Alton, \textit{et al. }(NuTeV Collaboration), Phys. Rev.
\textbf{D64}, 012002 (2001).

\bibitem {lepewwg}LEP/SLD Electroweak Working Group, hep-ex/0111221, with
updates from M. Gr\"{u}newald (private communication) for fits without
neutrino data and results posted at http://lepewwg.web.cern.ch/LEPEWWG/.

\bibitem {miller}G.A. Miller and A.W. Thomas, hep-ex/0204007, April 2002.

\bibitem {max prd}M. Goncharov, \textit{et al.} (NuTeV Collaboration), Phys.
Rev. \textbf{D64}, 112006 (2001).

\bibitem {nutev rapcom}G.P. Zeller, \textit{et al.}(NuTeV Collaboration),
hep-ex/0203004, March 2002.

\bibitem {sather}E. Sather, Phys. Lett. \textbf{B274}, 433 (1992).

\bibitem {rodionov}E.N. Rodionov, A.W. White, and J.T. Londergan, Mod. Phys.
Lett., \textbf{A} 9, 1799 (1994).

\bibitem {cao}F. Cao and A.I. Signal, Phys. Rev. \textbf{C62}, 015203 (2000).

\bibitem {langacker}P. Langacker, \textit{et al.}, Rev. Mod. Phys.
\textbf{64}, 87 (1991).

\bibitem {babu}K.S. Babu and J.C. Pati, hep-ph/0203029, March 2002.

\bibitem {barshay}S. Barshay and G. Kreyerhoff, hep-ph/0203054, April 2002.

\bibitem {giunti}C. Giunti and M. Laveder, hep-ph/0202152, Feb. 2002.

\bibitem {avva}S. Avvakumov, \textit{et al.} (NuTeV Collaboration),
hep-ex/0203018, March 2002.

\bibitem {ma-1}E. Ma and D.P. Roy, hep-ph/0111385, Dec. 2001.

\bibitem {ma-2}E. Ma, hep-ph/0112232, April 2002.
\end{thebibliography}
\end{document}